\documentclass[universe,article,accept,moreauthors,pdftex]{./Definitions/mdpi} 
\usepackage{caption}
\usepackage{subcaption}
\usepackage{subcaption}
\usepackage{graphicx}
\firstpage{1} 
\pubvolume{xx}
\issuenum{1}
\articlenumber{5}
\pubyear{2019}
\copyrightyear{2019}
\history{Received: date; Accepted: date; Published: date}

\pdfoutput=1

\Title{Nuclear physics at the energy frontier: recent heavy ion results from the perspective of the Electron Ion Collider}

\Author{Astrid Morreale $^{1,2}$\orcidA{},*}

\AuthorNames{Astrid Morreale}

\address{%
$^{1}$ \quad Institute for Globally Distributed Open Research, Paris France ; astrid.morreale@igdore.org\\
$^{2}$ \quad Center for Frontiers in Nuclear Science, New York, USA  ; astrid.morreale@gmail.com}

\corres{Correspondence: astrid.morreale@igdore.org}

\abstract{
Quarks and gluons are the fundamental constituents of nucleons. Their interactions rather than their mass, is responsible for $99\%$ of the
 mass of all visible matter in the universe. Measuring the fundamental properties of matter has had a large impact on our understanding of the nucleon structure and it has given us decades of  research and technological innovation. Despite the large number of discoveries made, many fundamental questions remain open and in need of a new and more precise generation of measurements.   The future Electron Ion Collider (EIC) will be a machine dedicated to hadron structure research. It will study the content of protons and neutrons in a largely unexplored regime in which gluons are expected to dominate and eventually saturate. 
While the EIC will be the machine of choice to quantify this regime, recent surprising results from the heavy ion community begin to exhibit similar signatures as those expected from a regime dominated by gluons.  Many  of the heavy ion results that will be discussed in this document  highlight the kinematic limitations of hadron-hadron and hadron-nucleus collisions. The reliability of using as a reference proton-proton (pp) and proton-Nucleus(pA) collisions to quantify and disentangle vacuum and Cold Nuclear Matter (CNM) effects from a  Quark Gluon Plasma  (QGP) may be under question. An selection of relevant pp and pA results which highlight the need of an EIC will be presented.}  

\keyword{QGP; EIC; Gluon Saturation; nPDF; DIS; CNM}

\begin{document}

\section{Introduction}\label{Sec:Intro}
Quarks and gluons, collectively called partons, are the fundamental constituents of protons, neutrons, the atomic nucleus as well as other hadrons. Their interaction is governed by Quantum Chromodynamics (QCD). Understanding QCD and in particular the confinement of quarks and gluons inside hadrons is one of today’s greatest physics challenges. QCD is the theory of strong interactions and it is expected to describe building blocks of visible matter and their binding in nuclei. While QCD is a well established theory,  it contains elements that cannot be calculated and rely mostly on experimental input.\footnote{While lattice calculations address these problem directly, results emerging from the lattice typically require large time scales. The accuracy of the obtained results is largely correlated with the amount of computing power allocated to pursuing these calculations.} As of today, many fundamental aspects of the theory have not yet been quantified. These aspects include the quantified contribution of partons (and their interactions) to the proton spin, or the mechanisms that permits us to transition from  point-like to non-point-like physics.  
%
%

Since the discovery of quarks and gluons and the confirmation that they carried color and spin. QCD and related sub-fields have continuously given us discoveries. One of which is the Quark Gluon Plasma (QGP) formation. QGP was the discovery that there was a state of matter in which partons were no longer confined to the  boundaries of a hadron, but rather acted as free particles.
Evidence  of this new state of matter was observed in heavy-ion collisions at the Relativistic Heavy Ion Collider
(RHIC) with the discovery of a suppression of high transverse momentum hadrons, also called "jet quenching~\cite{Adcox:2001jp}. Jet quenching is attributed to a decrease of the energy of the hard partons created during the first stages of a high-energy heavy-ion collisions.  
The formation of the QGP is now understood as being responsible for this loss of energy via interactions with its constituting hot and dense medium. Since its discovery we have learned many of the interesting properties governing the QGP:
\begin{itemize}
\item The QGP behaves as a near-ideal Fermi liquid (almost no frictional resistance or viscosity)~\cite{Cao:2010wa}.

\item The mean free path of partons in the QGP is comparable to inter-particle spacing ~\cite{Shuryak:2008eq}.

\item Experimental evidence points towards collective motion of particles during the QGP expansion~\cite{Adler:2002pu}.

\end{itemize}

\begin{figure}[h]
\begin{center}
\includegraphics[scale=2.0]{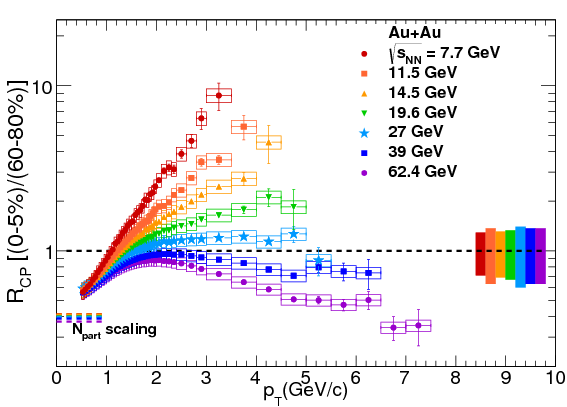}
\caption{Nuclear modification factor (R$_{CP}$) of high-$p_{T}$ hadrons produced in central collisions relative to those produced in peripheral collisions. A QGP onset is observed at collision energies $\sqrt(s_{NN})>$ 30$GeV$ while an enhancement is observed at lower energies~\cite{Adamczyk:2017nof}.}\label{Fig:RcpSTAR}
\end{center}
\end{figure}

While more precision measurements are needed,  some revealing information has been obtained regarding the QGP onset~\cite{Adamczyk:2017nof} as it is illustrated in Fig.~\ref{Fig:RcpSTAR} from the STAR experiment. This figure illustrates a classic QGP measurement:  particle suppression in heavy ion collisions observed via the central-to-peripheral nuclear modification factor ratio R$_{cp}$, as a function of transverse momentum and center of mass collision energy per nucleon-nucleon collision ($\sqrt(s_{NN})$). A smooth transition is seen as a function of $\sqrt(s_{NN})$ between enhancement and particle suppression, the latter a signature of the presence of a QGP.

Despite the plethora of information we have obtained regarding the QGP, many questions remain open. As an example we list: (1) How precisely  does the plasma acquire its Fermi like fluid characteristics  (2) What are the processes in which color-charged quarks and gluons and colorless jets interact with a nuclear medium  (3) Is there a smooth transition for the physics involved in small systems to that of large systems and (4) when does one transition from a regime of partons to a regime in which gluons dominate.

Indeed, recent puzzling results from proton-proton (pp) and proton-ion (pA) collisions seem to insist we address the above.


\section{A new physics regime}

\begin{figure}[h]
\begin{center}
\includegraphics[scale=0.21]{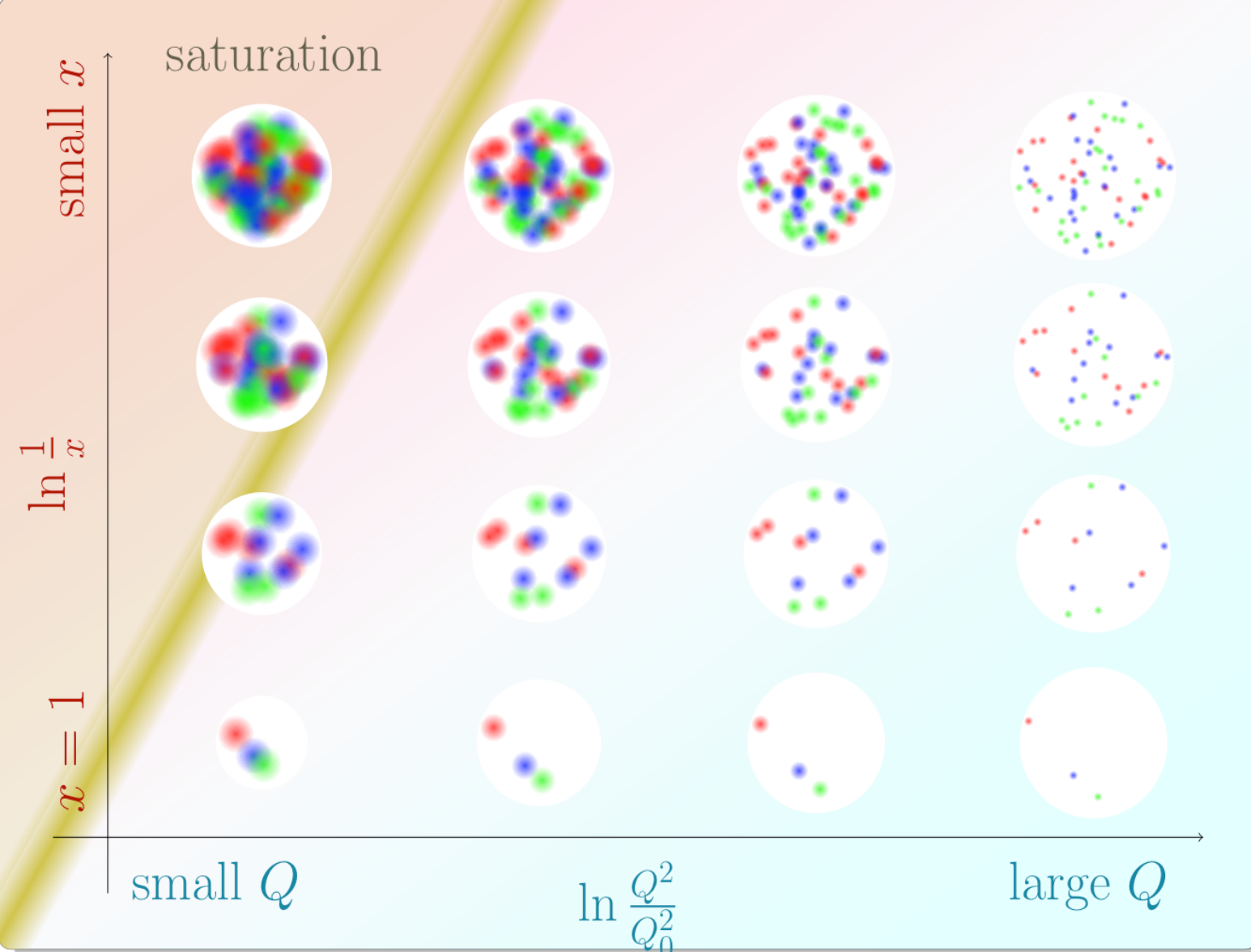}\includegraphics[scale=0.5]{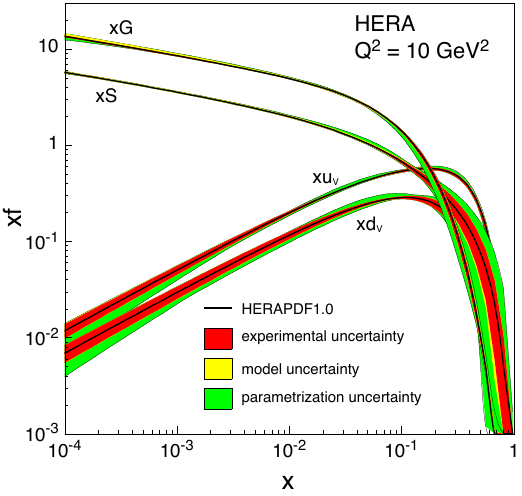}
\caption{Left: The QCD landscape, the horizontal axis  $Q^{2}$ represents the resolution of the probe while the Y axis ($ln(1/x)$) is related to the parton density. Figure from~\cite{Aschenauer:2017jsk}. Right: Parton distribution functions in the proton plotted as functions of Bjorken $x$. Figure from~\cite{Accardi:2012qut}}\label{Fig:QCDLandscape}
\end{center}
\end{figure}

The interaction between partons is usually described as a function of at least two quantities: the momentum fraction $x$ of the parent nucleon carried by the partons under consideration and the energy/length scale $Q^{2}$ at which the interaction between partons is probed. These two quantities allow one to identify several regimes for QCD, constituting what one calls the QCD landscape and illustrated on Fig.~\ref{Fig:QCDLandscape} (left). For a given $Q^{2}$ as we decrease towards smaller values of $x$ the number of partons is increased. While for a given $x$ and as we decrease towards smaller values of $Q^{2}$ (reduce the resolution) the size of partons increase. Now if we vary our kinematics towards small $x$ and small $Q$ one enters a regime characterized by a large number of partons (gluons rather), with overlapping wave functions. This is the phenomena that is known as  gluon saturation. 

For large values of  $Q^{2}$, the coupling constant $\alpha_{s}$ is small and one expects scattering directly from point-like bare color charges. pQCD can be then used to reliably predict the hard scattering of partons. For small values of  $Q^{2}$ , in a regime relevant for the description of nucleons and nuclei, one probes longer length scales making QCD non perturbative and very little is thus calculable.  For these small values of $Q^{2}$ the content of the nucleon in terms of partons is parameterized using parton distribution functions (PDF) and more recently Generalized Parton Distribution functions (GPD)~\cite{Ji:2004gf}. Parametrization of PDFs typically requires experimental input (or direct calculations on the  lattice.) For a given value of  $Q^{2}$ and decreasing values of $x$ the density of gluons in the nucleon increases very rapidly (see Fig.~\ref{Fig:QCDLandscape} right).  Yet for small enough  values of $x$, and large enough values of $Q^{2}$ for $\alpha_{s}$  to be considered small, it is expected that this increase eventually saturates, giving rise to a new regime characterized by weakly-coupled but highly correlated gluon matter called Color Glass Condensate (CGC).

 A variety of recent Large Hadron Collider (LHC) results indicate that small systems such as pp and pA exhibit signatures typically expected in larger heavy-ion systems (AA collisions) and  resulting from the presence of a QGP. A variety of theories exist which aim at providing explanation to these results some which include (1) presence of a QGP already in these small systems and (2) universal properties of all nuclei (small and large) in a gluon saturation regime. The first of these explanations requires a careful disentangling of the initial state effects. This is not a trivial task since this is usually achieved using these same small systems as a reference. The second explanation can be tested -with coarse precision and large uncertainties-  at current colliders. It is clear that a new generation of results such as those that will be performed at the Electron Ion Collider ( EIC) will be extremely important to help quantify initial state effects with better precision than what is currently achievable. Furthermore EIC measurements will be pivotal to precisely pin-down the presence of new physics regimes ie. gluon saturation.

\section{The Electron Ion Collider}\label{Sec:EIC}

\begin{figure}[H]
\centering
\includegraphics[scale=0.7]{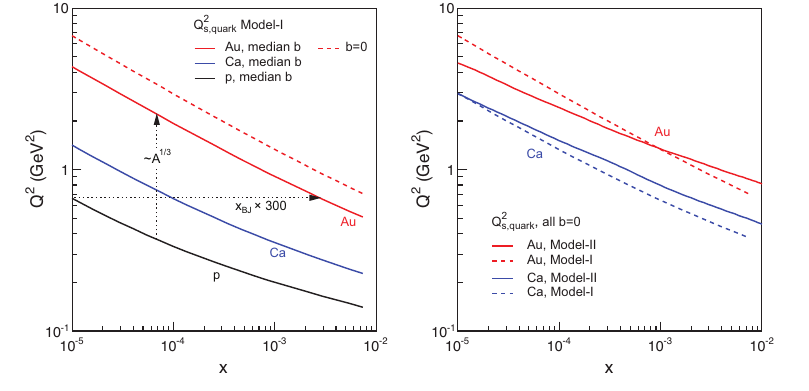}
\caption{Theoretical expectations for the saturation scale as a function of Bjorken $x$ for the
proton along with Ca and Au nuclei. $Q^{2}_{s}\sim 7 GeV^{2}$ is reached at $x=10^{-5}$ in e-p collisions at a $\sqrt{s}\sim 1~TeV$ while in e-Au collisions, only $\sqrt{s}\sim 60~GeV$ is needed to achieve comparable gluon density and the same saturation scale. Figure from~\cite{Accardi:2012qut}}\label{Fig:satScale}
\end{figure}

One of the goals of the lepton-ion (eA) program at an EIC is to unveil the collective behavior of densely packed gluons under conditions where their self-interactions dominate.~ 
 With its high luminosity and detector acceptance, as well as its span of available collision energies and ion species, the EIC will probe the confined motion as well as the spatial distributions of quarks and gluons inside a nucleus at one tenth of a femtometer resolution. The EIC will be able to detect soft gluons whose energy in the rest frame of the nucleus is less than one tenth of the average binding energy needed to hold the nucleons together to form the nucleus~\cite{Accardi:2012qut}.
Thanks to eA collisions with large nuclei, the EIC will reach the saturation regime faster than with ep collisions at similar cms energies (Fig.~\ref{Fig:satScale}). This is due to the $x$ and mass number (A) dependence of the saturation scale  $Q_{s}$ which goes like:$$Q^{2}_{s}(x)\sim A^{1/3}(1/x)^{\lambda}$$  The EIC will investigate the onset of saturation,  explore its properties and reveal its dynamical behavior. It will also  provide a kinematically well defined reference to quantify cold nuclear matter effects. For completeness it is noted that a similar accelerator proposal (LHeC) with complementary kinematic coverage and physics programe is being evaluated by the European Strategy for Particle Physics~\cite{Bruning:2652313,Klein:2018rhq}.

The EIC is considered a key component for the future nuclear physics program in the US and as such is among the key recommendations of the Nuclear Science Advisory Committee (NSAC) Long Range Plan from 2015. It has further received a positive and encouraging report from the National Academy of Sciences~\cite{NSAC}. 
\subsection{EIC requirements}
\begin{itemize}
\item Large luminosity ($10^{33}-10^{34}cm^{-2}s^{-1}$) 
\item Center of mass energy (30-140) GeV
\item Hadron and electron beams with high longitudinal spin polarization 
\item Ion beams from D to the heaviest sable nuclei
\item Large detector acceptance, in particular for small angle scattered hadrons
\item large detector acceptance, in particular for small angle scattered hadrons
\item[]  Optimized high luminosity and high acceptance running modes
\end{itemize}
\subsection{EIC designs}
The eRHIC design is based on an upgrade to the Relativistic Heavy Ion Collider (RHIC) located at Brookhaven National Laboratory (BNL) in  New York.

\begin{itemize}
\item New electron injector
\item 5-18 GeV electron energy
\item Heavy ions up to 100 $GeV/$u
\item $\sqrt{s}$: 20-140 GeV
\item Peak luminosity ~$\sim~0.4x10^{34}$cm$^{-2}$s$^{-1}/$A base design
\item $1.0x10^{34} cm^{-2}s^{-1}$/A with strong cooling
\end{itemize}

The JLEIC design is based on an upgrade to the Continuous Electron Beam Accelerator Facility (CEBAF) located at the Jefferson Laboratory in Virginia.

\begin{itemize}
\item New hadron injector
\item New figure eight collider configuration
\item 3-12 GeV electron energy
\item Heavy ions up to 80 GeV/u that could be upgraded to 160 GeV/u
\item $\sqrt{s}$: 20-100 GeV that  could be upgraded to 140 GeV
\item average luminosity per run $\sim 10^{34} cm^{-2}s^{-1}$/A
\end{itemize}

Both designs have science cases by themselves which require a robust integration with detector designs. An ongoing "Generic Detector for an EIC" research and development peer reviewed program is funded by the United States Department of Energy. Thanks to these funds an active effort exists in which a variety of detector designs and technologies which meet EIC requirements are being explored and tested.  Two such examples are cited: the BeAST and JELIC detector R$\&$D efforts. See~\cite{EICRandD} for a complete list of these programs.

\section{Physics at the energy frontier: selection of recent results}\label{Sec:Physics}
A short selection of unexpected heavy ion results measured at the LHC is presented. The presence of a mini QGP in small hadronic system~\cite{Zhang:2013oca,Park:2016jap}, has been proposed as an explanation. Other physics mechanisms that do not involve QGP formation have also  been proposed including the existence of a gluon saturation regime~\cite{Ma:2018bax,Mace:2018yvl}. For an interesting review on the subject see ~\cite{Dusling:2015gta}.
The selection presented hereafter will highlight the need to better understand small colliding systems if we are to quantify correctly QGP phenomena. The ep and eA collisions of the EIC  will undoubtly contribute to an in-depth understanding of these observations.

\subsection{QGP onset and strangeness enhancement}
\begin{figure}[H]
\centering
\includegraphics[scale=0.5]{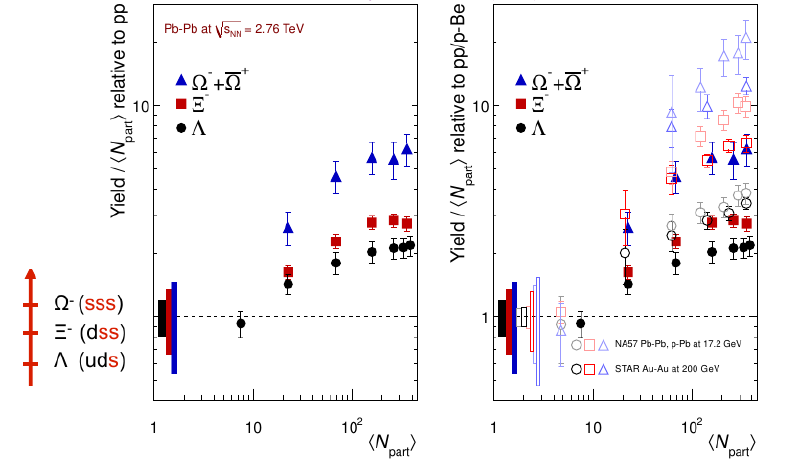}\includegraphics[scale=0.5]{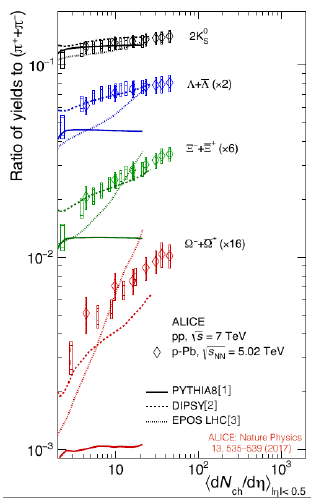}
\caption{Left and middle: Ratio of strange yields in PbPb collisions with respect to pp collisions as a function of participants. As it is observed in the figure, an enhancement with respect to pp is observed which is larger for  $\Omega$ (sss) than for $\Xi^{-}$ (dss) and $\Lambda$ (uds).  Rightmost: Ratio of strange yields to $\pi^{+}+\pi^{-}$ in pp, p-Pb a a function of average particle multiplicity. A smooth transition is observed as a function of particle multiplicity connecting the small (pp) and larger (pPb) systems.
}\label{Fig:strangenessEnhancement}
\end{figure}

Strangeness enhancement was one of the first proposed signatures of the QGP~\cite{Rafelski:1982pu}  
The  QGP expectation was that strange particle yields would be enhanced with respect to their yield in pp collision and that the enhancement would follow a hierachy based on their strange quark contents, namely that a particle with three strange quarks would be more enhance than one with two, and even more than a particle with only one strange quark
As predicted, strangeness enhancement was observed in AA collisions at the SPS, RHIC and the LHC (left and middle panels of  Fig.~\ref{Fig:strangenessEnhancement})~\cite {ABELEV:2013zaa}. 

What is unexpected however, is the observation (Fig.~\ref{Fig:strangenessEnhancement} right) that  an enhancement of strange particles (K, $\Lambda$, $\Omega$) with respect to non strange yields (ie $\pi$)  is also visible in the most violent high multiplicity pp and p-Pb collisions \footnote{Multiplicity is the number of charged particles in the final state. In pPb and PbPb this quantity is  related to the centrality of the collision. }.

 The mechanisms responsible for the observed enhancement in these small systems might indicate that such system may not be relied upon to discern cold from hot nuclear effects. While more experimental insight is needed to interpret the
the observed enhancement, it has been proposed that the presence of a strong gluon field leading to the non-linear regime of gluon saturation~\cite{Accardi:2012qut} may explain these observations.

\subsection{Heavy flavor vs Multiplicity}
\begin{figure}[H]
\centering
\includegraphics[scale=0.45]{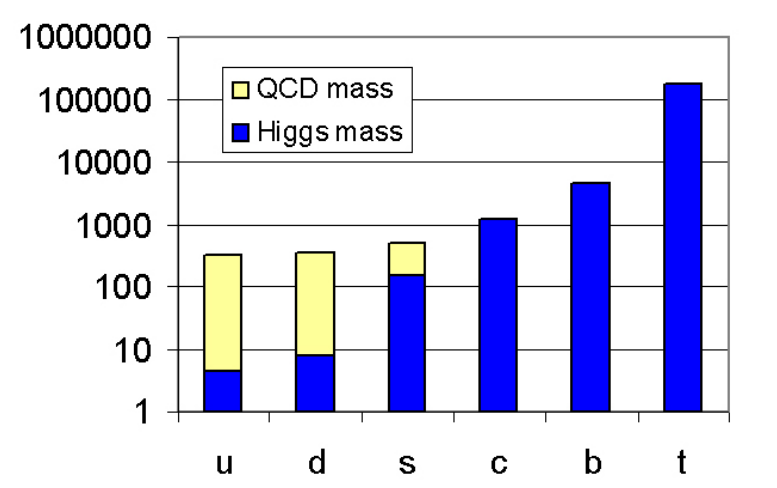}\includegraphics[scale=0.4]{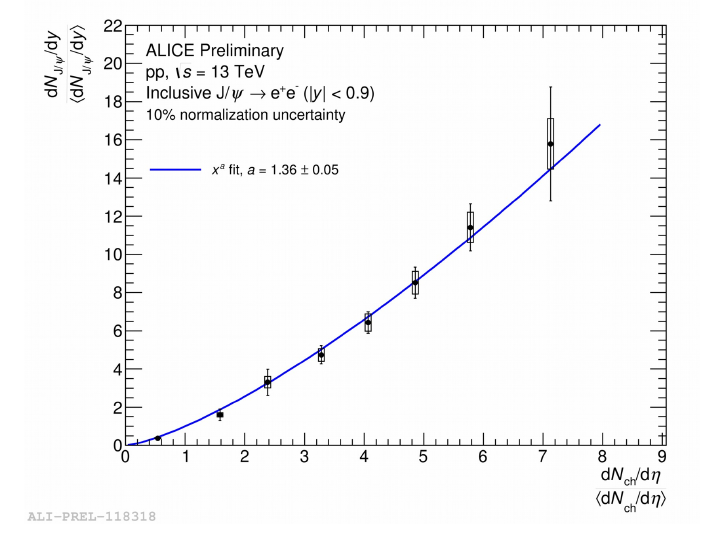}
\caption{Left: Masses of the six quark flavors. The masses generated by electroweak symmetry breaking (current quark masses) are shown in blue; the additional masses of the light quark flavors generated by spontaneous chiral symmetry breaking in QCD (constituent quark masses) are shown in yellow~\cite{Muller:2004kk}. Right: Relative J/$\psi$ production yields as a function of the relative number of charged particles per unit of rapidity. The blue line  corresponds it of a power law function to the data.}\label{Fig:HeavyVsLight}
\end{figure}

Heavy flavor probes are ideal to test QGP properties.  The contribution of the QCD vacuum condensate to the masses for the three light quark flavors (u, d, s) considerably exceeds the mass believed to be generated by the Higgs fields. Charm and beauty masses on the other hand, are not expected to be affected by this QCD vacuum (Fig.~\ref{Fig:HeavyVsLight} left) making them ideal probes of the QGP. 
The mass of the heavy quark itself provides the hard scale for pQCD calculations. This is in contrast to light quarks which often have to rely on the $p_{T}$ of the final state hadron. In addition low $p_{T}$ production of charmonia at forward rapidity (where smaller values of $x$ can be reached) is expected to be sensitive to gluon saturation.

 Recent results from the ALICE experiment at the LHC show an event activity dependence of inclusive $J/\psi$ and D mesons. The relative charmonium production yield as a function of the per-event relative charged particle multiplicity shows an increase that is faster than linear in pp collisions (Fig.~\ref{Fig:HeavyVsLight}right).  

\begin{figure}[H]
\centering
\includegraphics[scale=0.38]{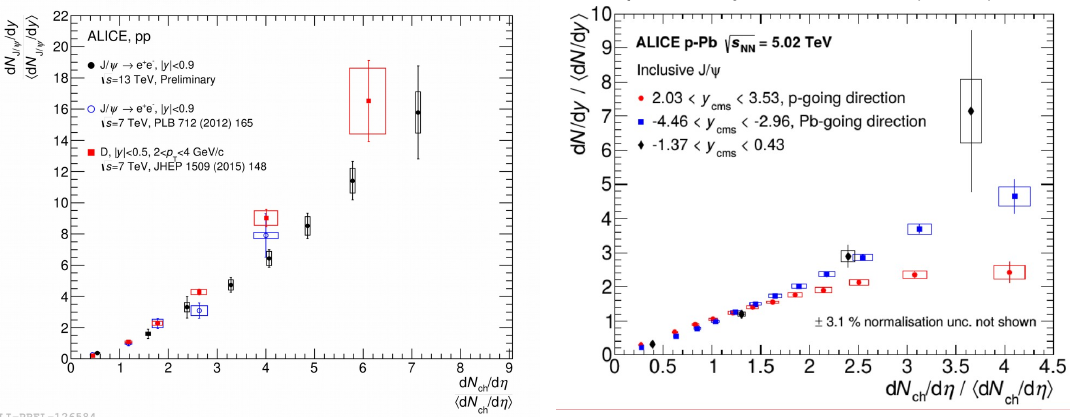}\includegraphics[scale=0.26]{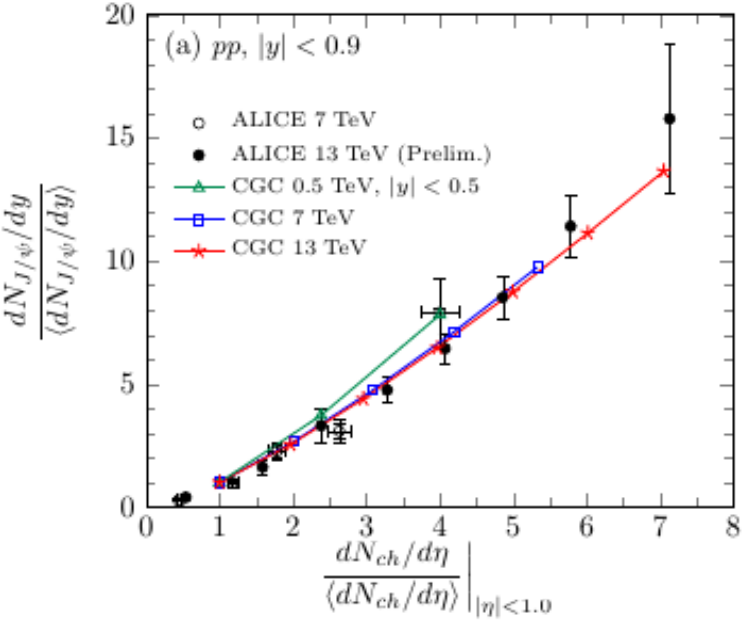}
\caption{Left: average inclusive $J/\Psi$ (closed black and open blue markers), D meson (red closed markers) dependence on charged particle multiplicity  in pp collisions and central rapidity. Middle: inclusive forward and backward rapidity $J/\psi$'s dependence on charged particle multiplicity  in p-Pb collisions~\cite{Adamova:2017uhu}. Right: CGC comparisons~\cite{Ma:2018bax} to recent  $J/\psi$  multiplicity results in pp collisions.}\label{Fig:MultiplicityJpsiAndD}
\end{figure}

Figure ~\ref{Fig:MultiplicityJpsiAndD} (middle)  shows a similar measurement performed in pPb collisions a negative (Pb-going side), mid and forward rapidity (p-going side). The positive rapidity measurement corresponds to small $x$ values ($\sim10^{-5}$) , a range in which gluon saturation may be present. The observation of similar charged particle multiplicity dependence (Fig.~\ref{Fig:MultiplicityJpsiAndD} left) for both open and hidden charm points that hadronization may be of lesser importance.

 One plausible physics explanation for the previous results is the existence of  a gluon saturation regime as it has been discussed in the introduction of this document. In Fig.~\ref{Fig:MultiplicityJpsiAndD} (right) a  CCG~\cite{Ma:2018bax} which includes gluon saturation effects  is compared to ALICE measurements in pp collisions. The calculation describes the data.


\subsection{Hydrodynamic flow}
One of the properties of the QGP is that it behaves like a perfect fluid with nearly zero viscosity. This near zero viscosity has been quantified by the correlated momentum anisotropies among the particles produced in the heavy collisions which result from a common velocity field pattern. This pattern is now identified as collective  flow~\cite{Heinz:2013th}. Among the flow phenomena, two types are highlighted in this document: (1) Radial flow which typically affects the shape of low $p_{T}$ spectra and (2) Elliptic Flow  $v_{2}$: which is the second coefficient of the Fourier decomposition of particle's momentum azimuthal distributions. This decomposition  quantifies the anisotropic particle density which emerges from two nuclei interacting in semi central collisions. A non zero $v_{2}$  implies early thermalization of the medium and it is considered a signature of the QGP.

\begin{figure}[H]
\centering
\includegraphics[scale=0.2]{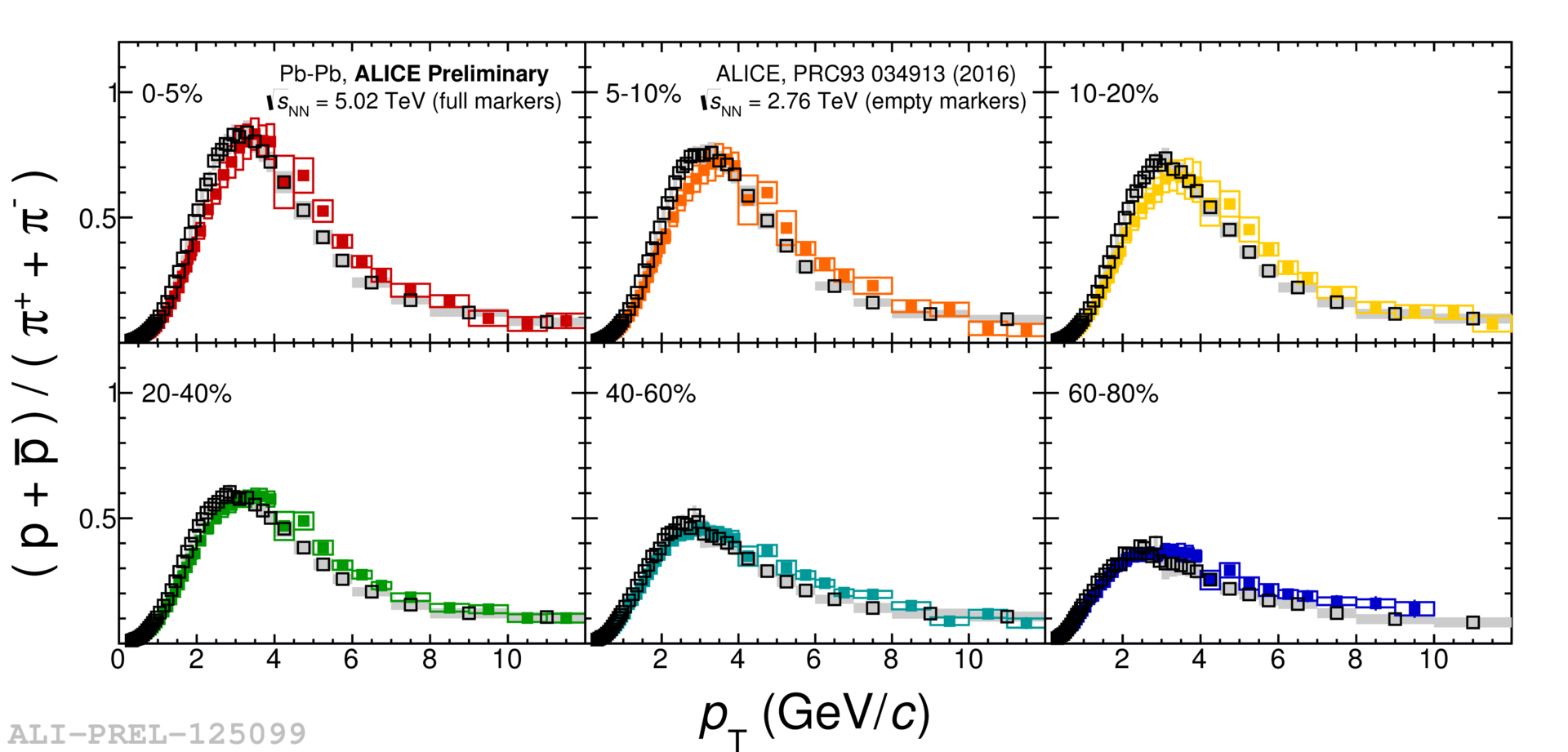}
\caption{Proton to pion ratio in PbPb collisions as a function of $p_{T}$ at two $\sqrt{s_{NN}}$ and six centrality classes. }\label{Fig:baryonMesonPbPb}
\end{figure}

\begin{figure}[H]
\centering
\includegraphics[scale=0.2]{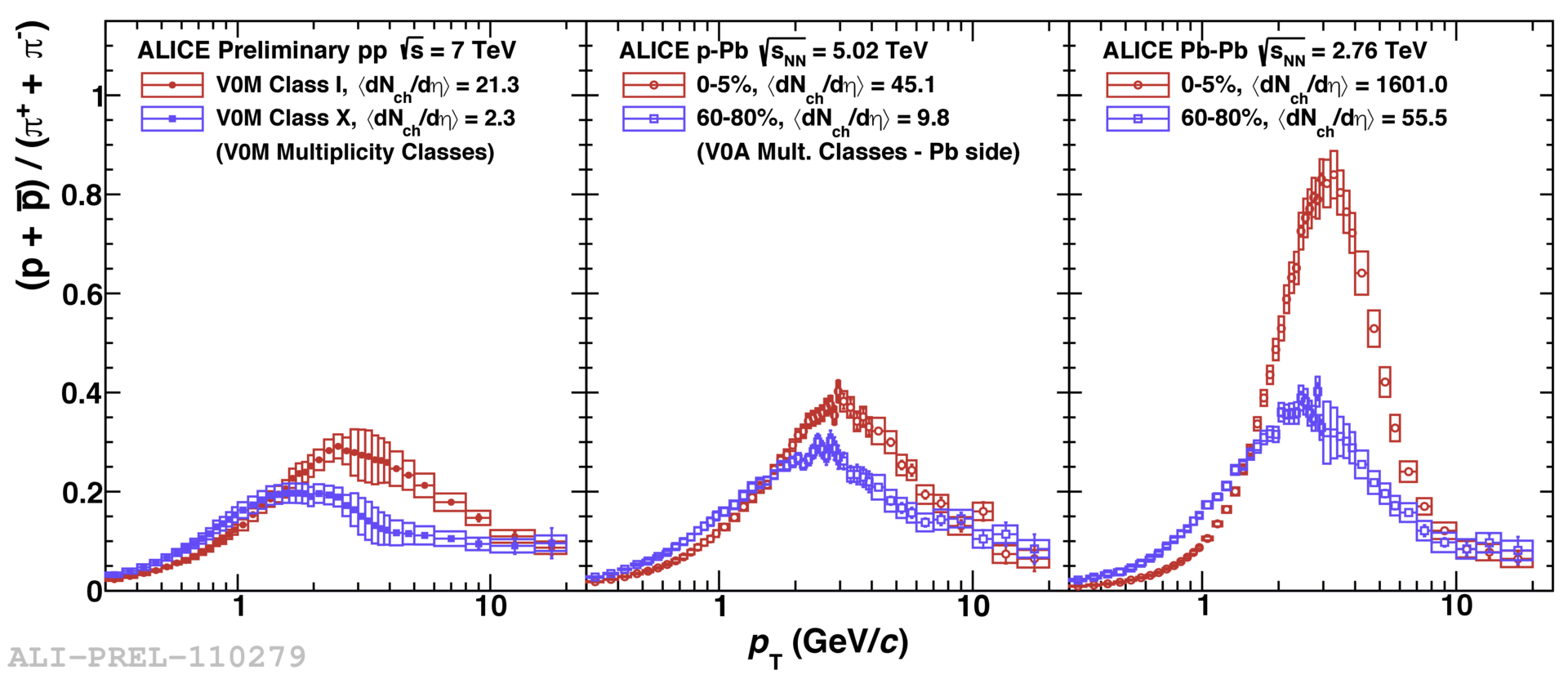}
\caption{Proton to pion ratio as a function of $p_{T}$ in pp (left), pPb (middle) and PbPb (right) collisions. The measurements are classified as a function of charged particle multiplicity.}\label{Fig:baryonMesonPP}
\end{figure}

Baryon to meson ratios obtained in Pb-Pb collisions as shown in Fig.~\ref{Fig:baryonMesonPbPb} illustrates the effect of radial flow. 
Radial flow will push hadrons from low $p_{T}$  towards intermediate $p_{T}$. The effect is expected to be stronger for baryons than for mesons, resulting in a bump in the baryon-to-meson ratio (here proton-to-pions), which depends on the centrality of the collision. Until recently, this was well understood in heavy-ion collisions. What is unexpected however, is the observation of a similar effect in pp and pPb collisions as shown in Fig.~\ref{Fig:baryonMesonPP}. The results would naively imply that thermalization is occurring already in these small systems.

\begin{figure}[H]
\centering
\includegraphics[scale=0.25]{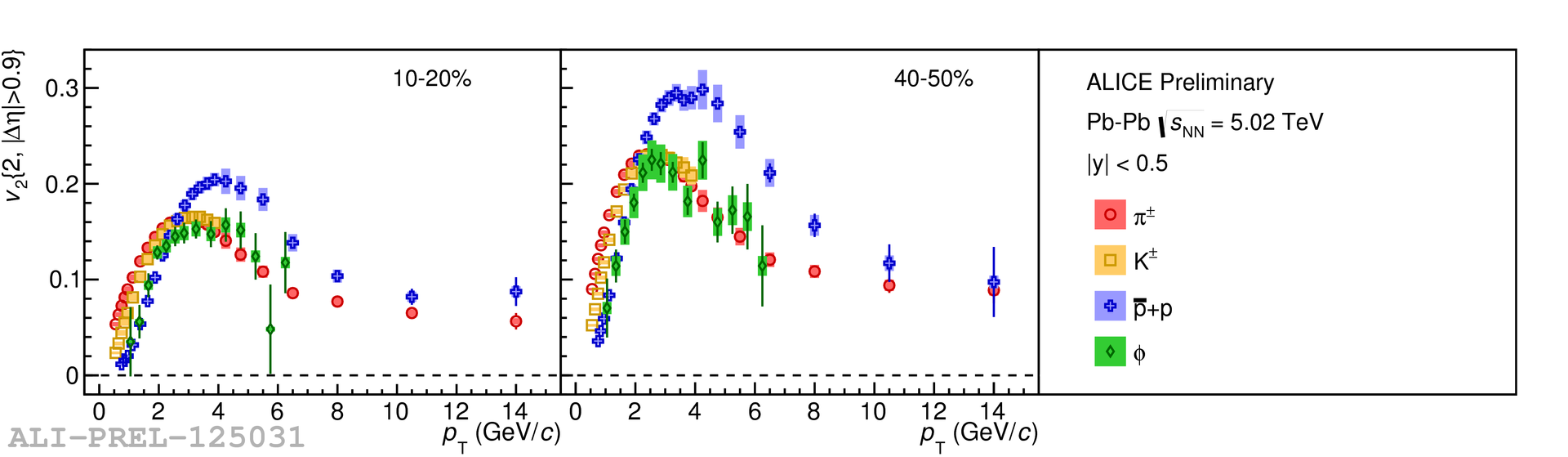}
\includegraphics[scale=0.15]{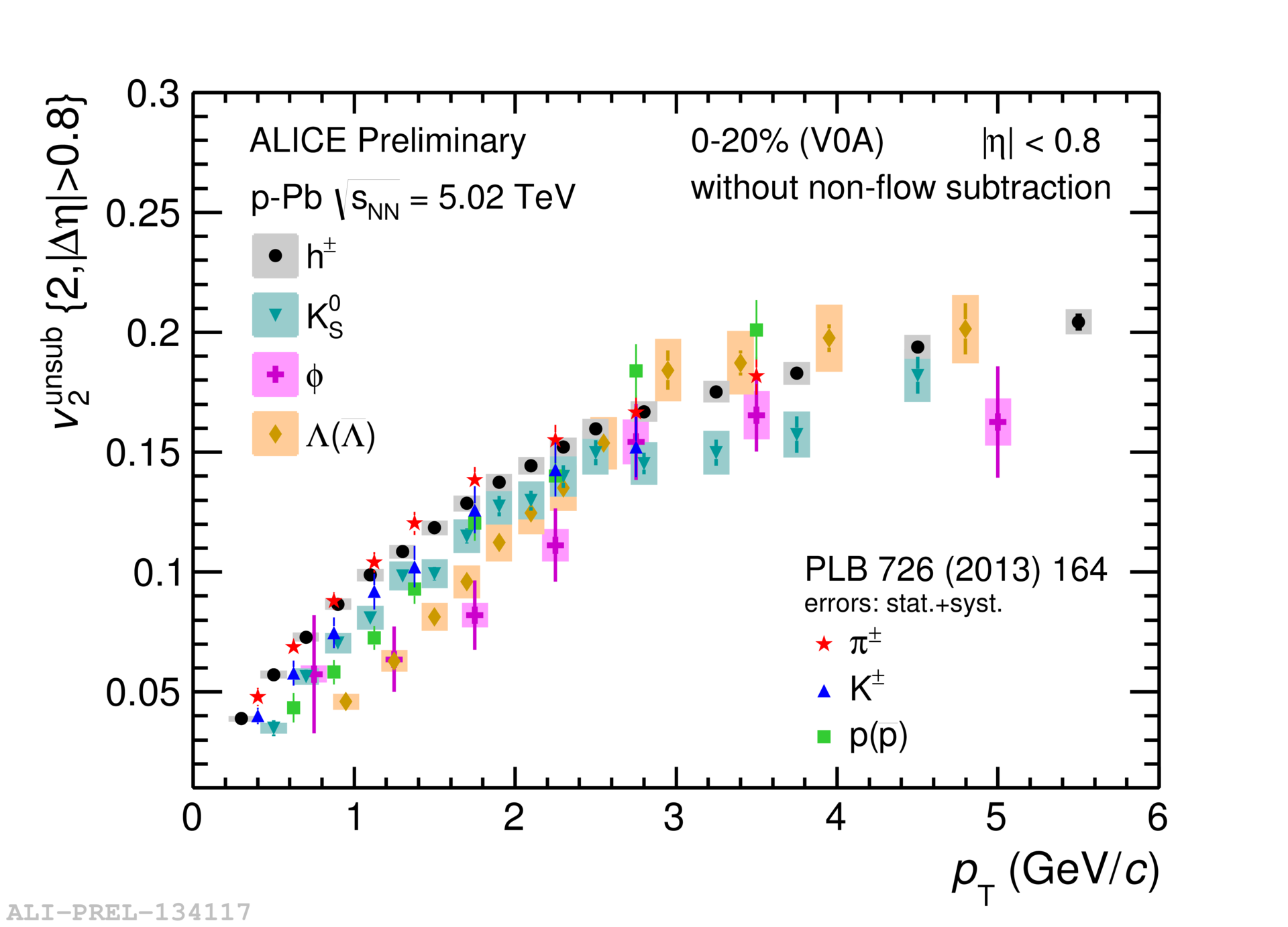}
\caption{Top figures: elliptic flow $v_{2}$  in PbPb collisions as a function of $p_{T}$ in two centrality classes and four particle species. Bottom figure: Elliptic flow $v_{2}$ in pPb collsions. }\label{Fig:flowPbPb}
\end{figure}

\begin{figure}[H]
\centering
\includegraphics[scale=0.7]{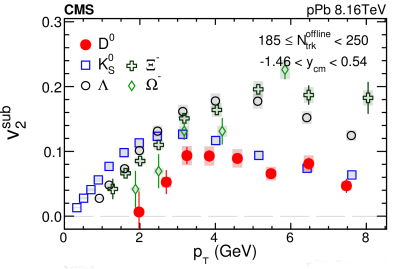}\includegraphics[scale=0.4]{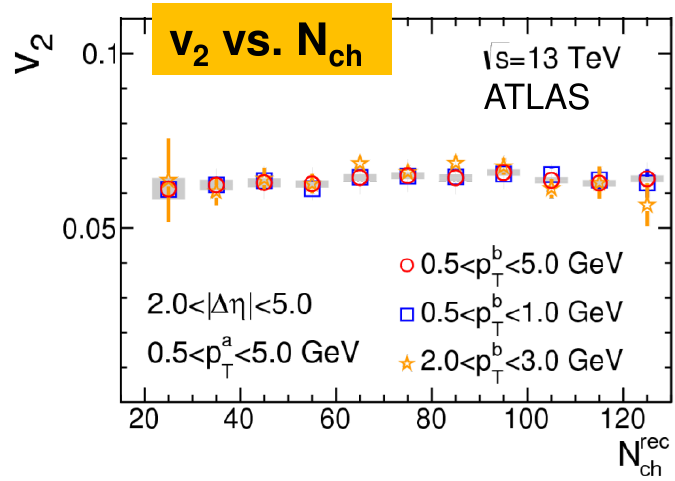}
\caption{Left panel: $v_{2}$ flow  as a function of $p_{T}$ of charm and strange hadrons in high-multiplicity pPb collisions at $\sqrt{s}_{pA}=8.16~TeV$ (CMS Collaboration~\cite{Sirunyan:2018toe}). Right panel: $v_{2}$ as a function of $p_{T}$ in pp collisions at  $\sqrt{s}=13 TeV$ (ATLAS Collaboration~\cite{Aad:2015gqa}.)}\label{Fig:flowHF}
\end{figure}

Light meson flow $v_{2}$ results in PbPb and reported by the ALICE collaboration are shown in Fig.~\ref{Fig:flowPbPb} (top figures). At low $p_{T}$, as it is the case for many other flow results, the trend is understood as being consistent with a collective expansion within the QGP  and has been successfully explained by 
 hydrodynamic models~\cite{Hirano:2008hy}.  
At high $p_{T}$ constituent quark number scaling takes over (dressed quarks), all mesons fall together and baryons climb above by $\sim 1/3$. What is intriguing on the other hand is that similar signatures are observed in pPb (Fig.~\ref{Fig:flowPbPb} bottom and Fig.~\ref{Fig:flowHF} ) and pp collisions (Fig.~\ref{Fig:flowHF} bottom right ). Effects than can cause the current observations are either due to initial state effects (saturation), or final state effects (expansion and/or thermal equilibrium). More recently, quantum entanglement has been suggested as a possible explanation~\cite{Bellwied:2018gck} and well as double 
parton scattering coupled with the  elliptic gluon Wigner distributions~\cite{article}. This phenomena could be elucidated with a variety of probes at the  EIC's lepton-nucleon program, including diffractive measurements of dijet production.

\subsection{Nuclear modification factor and energy loss in the medium}
The nuclear modification factor $R_{AA}$ is an observable used to quantify the effect of the nuclear medium on particle production. $R_{AA}$ consists of measuring invariant spectra as a function of p$_{\rm T}$ of particles produced in heavy ion collisions and compared to reference data (pp) at the same energy and scaled by the number of binary collisions. $R_{AA}$  is defined as:
 $$R_{AA}=\frac{\rm AA}{\rm scaled~ pp} =\frac{d^2N_{\rm AA}/dp_{\rm T}dy}{<N_{\rm coll}>d^{2}N_{\rm pp}/dp_{\rm T}dy} $$
Values greater than unity would be an indication of production enhancement, while values less than unity will indicate particle suppression in the QGP. 

\begin{figure}[H]
\centering
\includegraphics[scale=0.2]{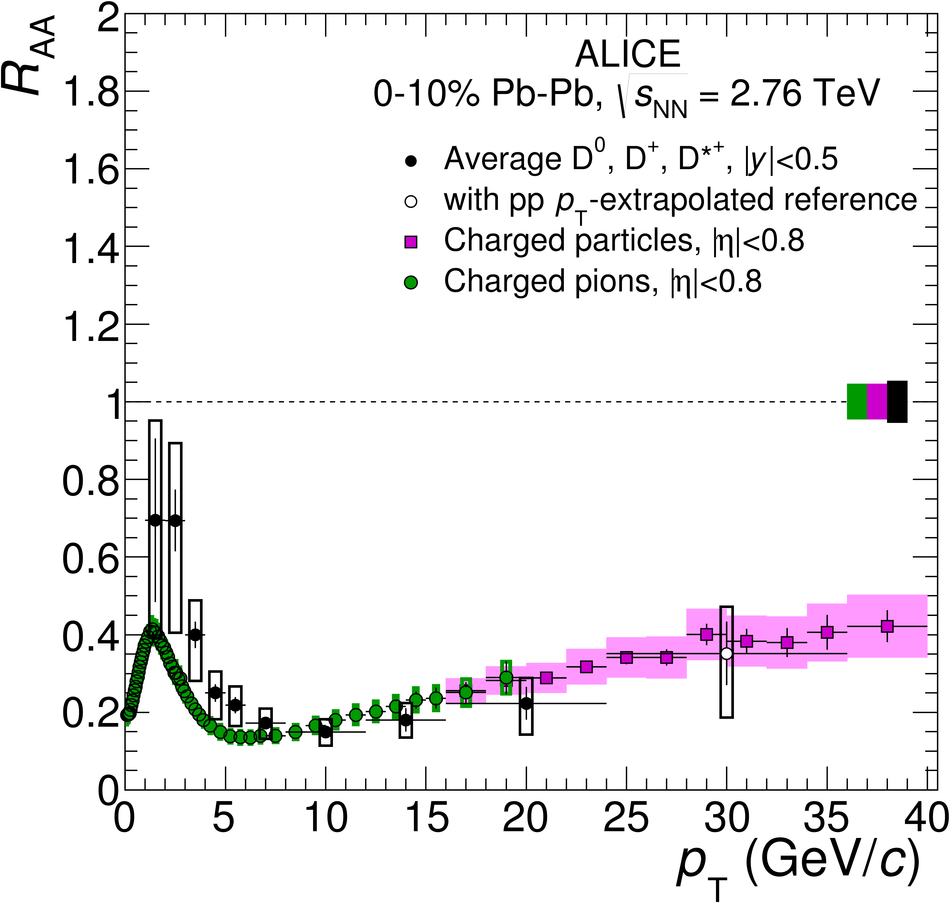}\includegraphics[scale=0.105]{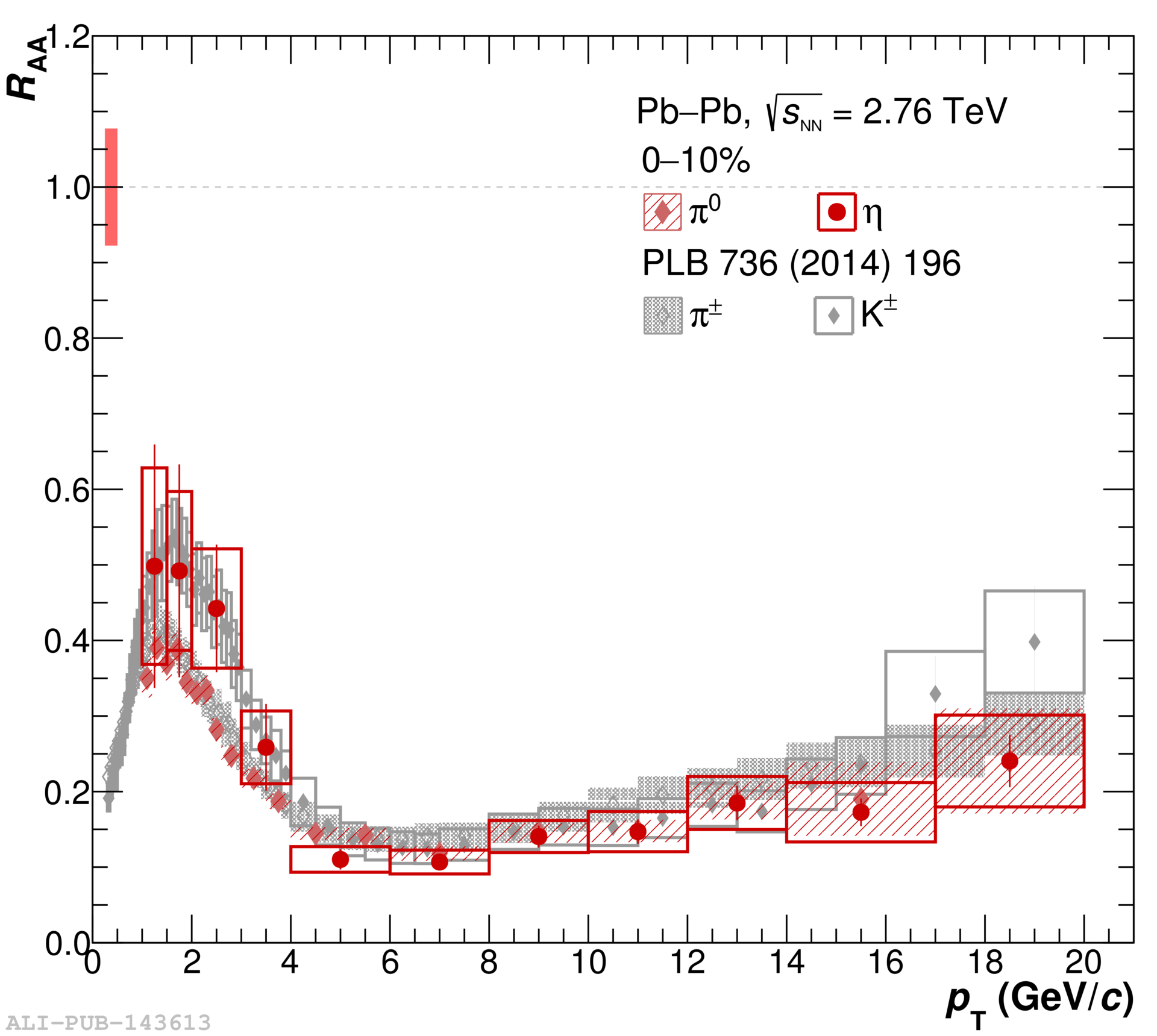}
\caption{Left: Prompt D-meson R$_{AA}$ as a function of p$_{T}$ compared to the nuclear modification factors of charged pions and charged particles in the 0-10$\%$ centrality class~\cite{Adam:2015sza}. Right: R$_{AA}$ of neutral and charged pions, kaons and eta meson ~\cite{Acharya:2018yhg}}\label{Fig:RAA}
\end{figure}

While partons are expected to loose energy when propagating through the dense QGP medium it is also expected that the amount of energy loss will depend on the parton type and the medium properties.  A large number of results such as those in Fig.~\ref{Fig:RAA} indicate that the amount of suppression observed in heavy ion collisions is irrelevant of particle mass (or quark content) at high enough p$_{\rm T}$. 

\begin{figure}[H]
\centering
\includegraphics[scale=0.4]{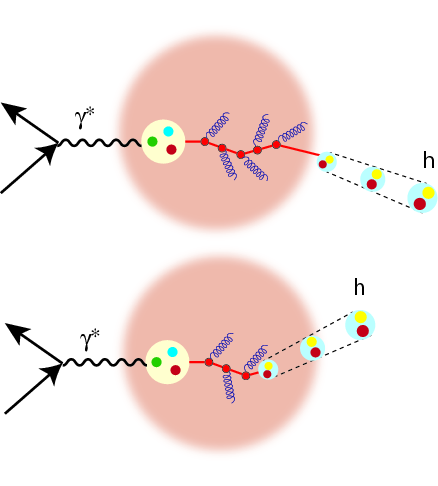}\includegraphics[scale=0.3]{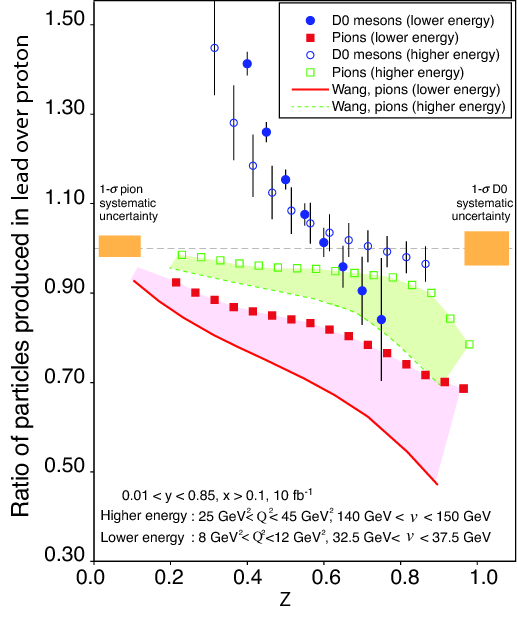}
\caption{{\bf Left:} Hadronization schematic illustrating the interaction of a parton moving through cold nuclear matter: the hadron is formed outside (top) or inside (bottom) the nucleus. {\bf Right:} Ratio of semi-inclusive cross section for producing a pion (red) composed of light quarks, and a D0 meson (blue) composed of heavy quarks in e-Lead collisions to e-deuteron collisions, plotted as function of z, the ratio of the momentum carried by the produced hadron to that of the virtual photon ($\gamma*$), as shown in the plots on the left. Figures and descriptions taken from~\cite{Accardi:2012qut}}\label{Fig:EnergyLossEIC}
\end{figure}

$R_{AA}$ results could largely benefit from independent measurements at the EIC.  Measurements such as those illustrated in Fig.~\ref{Fig:EnergyLossEIC} will study the response of the nuclear medium to a fast moving quark~\cite{Accardi:2012qut,Aschenauer:2017jsk} and allow proper understanding of hadronization mechanisms. 

\subsection{Nuclear parton distribution functions}

\begin{figure}[H]
\centering
\includegraphics[scale=0.4]{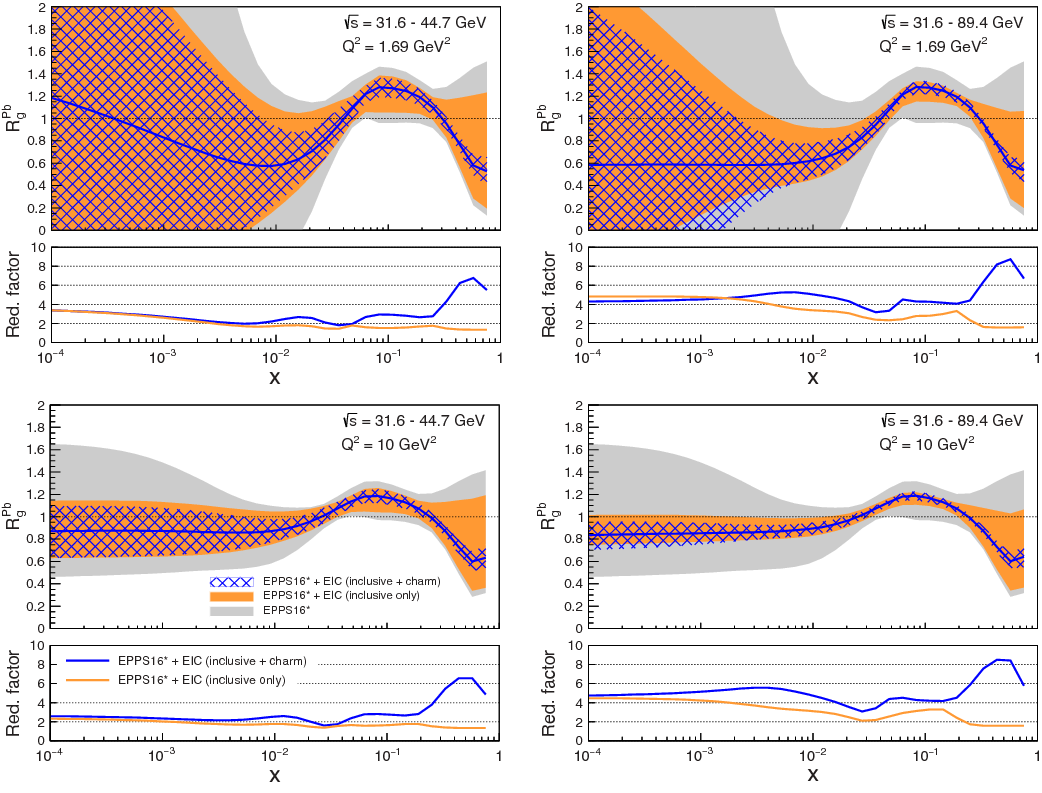}
\caption{{\bf Left:}The ratio R$^{Pb}_{g}$, from EPPS16$*$, of gluon distributions in a lead nucleus relative to the proton, for two different momentum transfers Q$^{2}$ possible at the EIC. The grey band represents the current theoretical uncertainty. The orange (blue hatched) band includes the EIC simulated inclusive (charm quark) reduced cross-section data. The lower panel in each plot shows the reduction factor in the uncertainty with respect to the baseline fit. Figures and details taken from ~\cite{Aschenauer:2017jsk}}\label{Fig:nPDFEIC}
\end{figure}

Finally, a careful evaluation of initial state effects such as nuclear modifications of Parton Distribution Functions (nPDFs) is also needed in order to correctly quantify hot nuclear effects present.

nPDFs refers to the difference observed between nuclear (bound nucleons) PDFs and and free nucleons PDFs (proton, neutron). The nuclear modification of PDFs is due to the interactions between partons from different nucleons. As such,  precise measurements of nPDFs are essential in order to understand cold nuclear matter effects that may be convoluted with current heavy ion results.
Fig.~\ref{Fig:nPDFEIC} illustrates (in grey) the uncertainty of gluons distributions in the lead nucleus which is rather large at both low and high $x$. Measurements that aim at at improving the precision on nPDF are proposed key measurements  of the EIC~\cite{Aschenauer:2017jsk}.
 
\section{Conclusions}
QCD studies have given us decades of discoveries. Many open questions remain on how does the transition from a small system to a dense system takes place: this information is needed to fully understand the properties of the QGP. The current document has given a selection of  results that may be better understood and quantified with a new generation of lepton-nucleon experiments at the EIC.

\vspace{6pt} 



\acknowledgments{The author thanks the Center for Frontiers on Nuclear Science for funding support which enabled promoting the topics outlined in this document.}



\externalbibliography{yes}
\bibliography{bibliography}



\end{document}